\documentclass[
  aps,prl,reprint,superscriptaddress,
  longbibliography,amsmath,amssymb,
  floatfix,nobibnotes
]{revtex4-2}

\ifdefined\pdfminorversion
\fi

\usepackage{graphicx}
\usepackage{bm}
\usepackage[T1]{fontenc}
\usepackage[hidelinks]{hyperref}

\newcommand{\wLO}{\omega_{\mathrm{LO}}}
\newcommand{\GaLO}{\Gamma^{\mathrm{LO}}_0}
\newcommand{\GeffLO}{\Gamma^{\mathrm{LO}}_{\mathrm{eff}}}
\newcommand{\Galloy}{\Gamma_{\mathrm{alloy}}}
\newcommand{\Giv}{\Gamma_{\mathrm{IV}}}
\newcommand{\thph}{\theta_{\mathrm{ph}}}
\newcommand{\thzero}{\theta_0}
\newcommand{\mstar}{m^*}
\newcommand{\ET}{E_{\mathrm{th}}}
\newcommand{\Ekin}{\mathcal{E}}
\newcommand{\Fld}{\mathcal{F}}
\newcommand{\kvec}{\bm{k}}
\newcommand{\lsp}{\ell_{\mathrm{sp}}}
\newcommand{\aEn}{\alpha^{*}}
\newcommand{\lLO}{\ell_{\mathrm{LO}}}
\newcommand{\prlhead}[1]{\textit{#1}---}

\begin{document}

\title{Disorder into Correlation: A Microscopic Mechanism for Sub-McIntyre Avalanche Noise}

\author{Shafat Shahnewaz}
\affiliation{Department of Electrical and Computer Engineering, University of Virginia, Charlottesville, Virginia 22903, USA}
\author{Hannaneh Karimi}
\affiliation{Department of Electrical and Computer Engineering, University of Virginia, Charlottesville, Virginia 22903, USA}
\author{Joe C. Campbell}
\affiliation{Department of Electrical and Computer Engineering, University of Virginia, Charlottesville, Virginia 22903, USA}
\author{Avik W. Ghosh}
\email{ag7rq@virginia.edu}
\affiliation{Department of Electrical and Computer Engineering, University of Virginia, Charlottesville, Virginia 22903, USA}
\affiliation{Department of Physics, University of Virginia, Charlottesville, Virginia 22903, USA}

\begin{abstract}
Avalanche multiplication is intrinsically stochastic, yet experiments reveal excess noise below the minimum predicted by McIntyre’s local-field theory. The microscopic origin of this noise suppression remains incompletely understood. We show that polar-optical-phonon emission, confined to a narrow forward cone, competes with elastic alloy scattering that diffuses field-collimated hot carriers out of the cone. Suppressed phonon cooling narrows the ionization path beyond the dead space and synchronizes successive ionizations into a dynamic staircase. This explains the low noise of Sb-based alloys and HgCdTe and makes alloy disorder a design parameter.
\end{abstract}

\maketitle

\prlhead{Introduction}Avalanche photodiodes (APDs) provide internal gain by accelerating photogenerated carriers until they impact ionize, lifting weak photocurrents above downstream electronic noise in optical receivers, LiDAR, and single-photon imaging~\cite{Tosi2014,Hadfield2023}. The avalanche is intrinsically statistical, and the variance of the multiplication $m$ produced by one injected carrier appears as the excess-noise factor
\begin{equation}
F(M)=\frac{\langle m^2\rangle}{\langle m\rangle^2}=1+\frac{\sigma_m^2}{M^2},
\label{eq:definition_F}
\end{equation}
where $M=\langle m\rangle$ is the mean gain. In McIntyre's local-field theory~\cite{McIntyre1966,McIntyre1972}, for electron injection,
\begin{equation}
F(M)=kM+(1-k)\left(2-\frac{1}{M}\right),
\label{eq:mcintyre}
\end{equation}
where $k=\beta/\alpha$ is the hole-to-electron ionization-coefficient ratio. Even for $k\to0$, the memoryless local model gives $F\to2$ at high gain. Reproducible noise below this curve therefore requires revisiting the ionization and avalanche statistics~\cite{Burgess1959}. A finite dead space introduces carrier history: after ionization, the carrier must regain threshold energy before ionizing again. Localization of successive ionizations can consequently reduce multiplication fluctuations.

Several heavily alloyed APD materials indeed fall below this bound. Electron-initiated Hg$_{1-x}$Cd$_{x}$Te APDs reach gains above $10^3$ with $F$ close to unity~\cite{Kinch2004,Derelle2009}, and Sb-based AlInAsSb and AlGaAsSb APDs show low excess noise that does not follow the constant-$k$ McIntyre curves~\cite{Jones2020,Karimi2025}. Monte Carlo studies indicate that strong alloy scattering and weak optical-phonon cooling are both relevant~\cite{Derelle2009,Karimi2025}, but the microscopic link between alloy disorder, phonon cooling, dead-space correlation, and sub-McIntyre noise has not been articulated.

\begin{figure*}[t]
\centering
\includegraphics[width=0.8\textwidth]{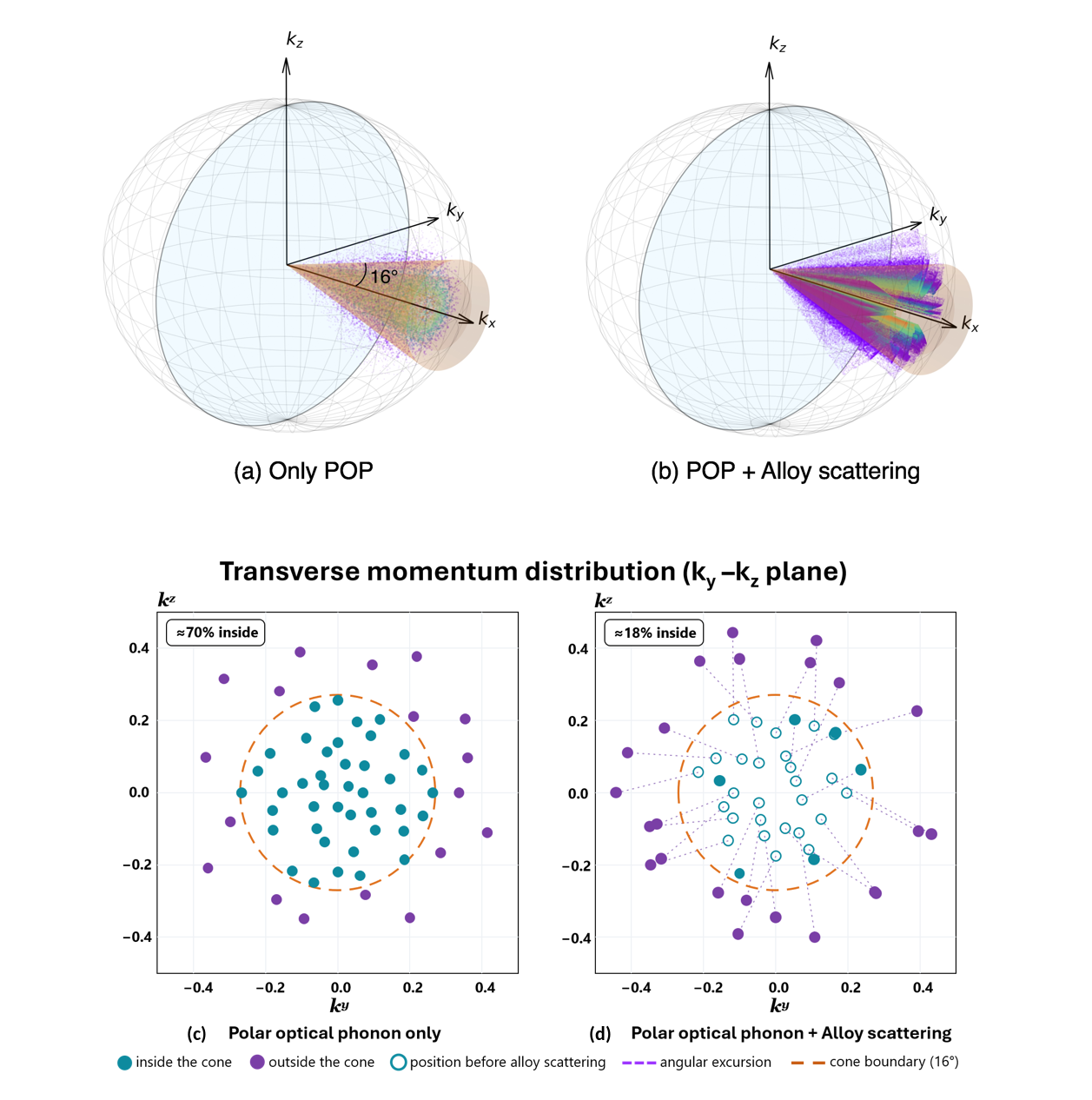}
\caption{Electron ensemble in $k$ space with carriers accelerated along $k_x$. (a)~POP emission only: carriers remain inside the forward Fr\"ohlich emission cone (half-angle $16^\circ$, shaded). (b)~POP plus elastic alloy scattering: alloy events rotate the momentum and spread the ensemble outside the cone. (c),(d)~Corresponding transverse $(k_y,k_z)$ distributions; the dashed circle is the $16^\circ$ cone boundary. Filled teal (purple) symbols lie inside (outside) the cone; in (d), open circles mark positions before alloy scattering and dotted lines the resulting angular excursions. In the device Monte Carlo of Al$_{0.7}$InAsSb, the in-cone fraction falls from ${\approx}70\%$ to ${\approx}18\%$ when alloy scattering is included.}
\label{fig:kspace}
\end{figure*}

In this Letter we propose that alloy disorder acts through an angular channel. Fr\"ohlich polar-optical-phonon (POP) scattering is forward weighted, with a matrix element proportional to $1/q^2$~\cite{Ridley1999,Lundstrom2000}. The field collimates hot carriers along its axis [Fig.~\ref{fig:kspace}(a)], while elastic alloy scattering rotates their momenta [Fig.~\ref{fig:kspace}(b)]. In the proposed picture, this angular redistribution suppresses effective POP cooling. After impact ionization, fewer cooling events during recovery shorten the stochastic tail of the ionization path. Multiplication then resembles a staircase APD~\cite{Capasso1983}, with steps generated dynamically by dead-space recovery rather than heterojunction offsets.

\prlhead{Dead space and spillover}The APD shot-noise spectral density is $S_I=2q(I_{\mathrm{ph}}+I_{\mathrm{dark}})M^2F(M)$, where $I_{\mathrm{ph}}$ and $I_{\mathrm{dark}}$ are the unity-gain photocurrent and dark current; the mean-square noise over bandwidth $\Delta f$ is $S_I\Delta f$. Reducing excess noise means reducing the relative variance of multiplication, not merely raising the mean gain. A carrier emerging from an ionization event is cold relative to the threshold energy $\ET$ and must travel an approximately deterministic distance
\begin{equation}
d\simeq \frac{\ET}{q\Fld}
\label{eq:deadspace}
\end{equation}
in a field $\Fld$ before it can ionize again~\cite{Saleh1990,Hayat1992}. The ionization path-length distribution therefore has no weight below $d$ and a peak shifted toward the recovery length. Beyond the dead space a carrier ionizes after a further random distance, described by the enabled (post-dead-space) ionization coefficient $\aEn$, with path-length density $\aEn e^{-\aEn(x-d)}$ for $x\ge d$~\cite{Ng2002}. We identify the post-dead-space spillover length $\lsp\equiv1/\aEn$ as the stochastic distance over which an energized carrier samples phonon cooling, alloy deflection, intervalley transfer, and impact ionization after reaching $\ET$. Our central hypothesis is
\begin{equation}
\begin{gathered}
\lsp=1/\aEn\lesssim d\\
\Longrightarrow\quad\text{correlated multiplication.}
\end{gathered}
\label{eq:centralcriterion}
\end{equation}
If $1/\aEn$ is shorter than $d$, successive ionizations are synchronized and the avalanche becomes staircase-like; if it is comparable to or larger than $d$, the statistics revert toward Eq.~\eqref{eq:mcintyre}. Equating mean path lengths relates $\aEn$ to the conventional local coefficient $\alpha$ through $1/\alpha=d+1/\aEn$~\cite{Ng2002}; $\alpha$ lumps the deterministic and stochastic parts of the path together, whereas $\aEn$ isolates the stochastic part.

\prlhead{Forward-weighted POP emission}For Fr\"ohlich coupling $|M(\bm{q})|^2\propto 1/q^2$~\cite{Ridley1999}. For emission of a phonon $\hbar\wLO$ that takes $\kvec\to\kvec'$ through an angle $\phi$, $q^2=k^2+k'^2-2kk'\cos\phi\simeq(k-k')^2+kk'\phi^2$ at small $\phi$, and for carrier energy $\Ekin\gg\hbar\wLO$ ($k'\simeq k$),
\begin{equation}
|M|^2 \propto \frac{1}{(k-k')^2+k^2\phi^2}.
\label{eq:smallangle_matrix}
\end{equation}
POP emission thus removes energy while leaving the carrier direction nearly unchanged. The denominator crossover angle is $\phi_{\mathrm{peak}}\simeq(k-k')/k\simeq\hbar\wLO/2\Ekin$, but the phase-space factor $\sin\phi$ broadens the cumulative emission probability, so an order-one containment cone scales as (see End Matter)
\begin{equation}
\thph^2 \sim \frac{\hbar\wLO}{\Ekin}.
\label{eq:phononcone}
\end{equation}
For Al$_{0.7}$InAsSb ($\hbar\wLO\approx24\,\mathrm{meV}$~\cite{Karimi2025}) this gives $\thph\approx16^\circ$ at $\Ekin=0.3\,\mathrm{eV}$, narrowing to ${\approx}6^\circ$ near the electron threshold $\ET\approx2\,\mathrm{eV}$~\cite{Karimi2025}; for GaAs ($\hbar\wLO\approx36\,\mathrm{meV}$) the same estimate gives ${\approx}20^\circ$ at 0.3~eV. Only the existence of a limited forward cone matters below; its numerical prefactor does not.

\prlhead{Field collimation versus alloy diffusion}The field also focuses the momentum distribution. With $\theta$ the angle between $\kvec$ and the field axis, $\hbar\dot{\kvec}=q\bm{\Fld}$ gives, to leading order, a collimating drift
\begin{equation}
\begin{aligned}
\dot\theta&=-A\sin\theta,\\
A&=\frac{q\Fld}{\hbar k}=\frac{q\Fld}{\sqrt{2\mstar\Ekin}}.
\end{aligned}
\label{eq:angular_drift}
\end{equation}
Alloy scattering adds angular diffusion with coefficient $D_\theta\simeq C_a\Galloy$, where $C_a$ is an order-unity factor set by the angular kernel. The angular distribution $f(\theta,t)$ then obeys the Fokker--Planck equation
\begin{equation}
\frac{\partial f}{\partial t}
=\frac{1}{\sin\theta}\frac{\partial}{\partial\theta}
\left[A\sin^2\theta\,f+D_\theta\sin\theta\,\frac{\partial f}{\partial\theta}\right],
\label{eq:fp_full}
\end{equation}
whose regular small-angle steady state (End Matter) is
\begin{equation}
f(\theta)\propto \exp\!\left(-\frac{\theta^2}{2\thzero^2}\right),
\qquad
\thzero^2=\frac{D_\theta}{A}.
\label{eq:theta0}
\end{equation}
The field narrows $\thzero$; alloy disorder widens it. The fraction of carriers inside the POP cone sets the phonon-suppression factor
\begin{equation}
R_{\mathrm{ph}}\equiv\frac{\GeffLO}{\GaLO}=P_{\mathrm{cone}}
=1-\exp\!\left(-\frac{\thph^2}{2\thzero^2}\right),
\label{eq:Rph}
\end{equation}
and in the alloy-broadened limit $\thzero\gg\thph$,
\begin{equation}
\begin{aligned}
R_{\mathrm{ph}}&\simeq\frac{A\thph^2}{2D_\theta}\\
&\propto\frac{q\Fld}{\Galloy\sqrt{2\mstar\Ekin}}\,
\frac{\hbar\wLO}{\Ekin}.
\end{aligned}
\label{eq:suppression_scaling}
\end{equation}
The phonon-limited regime has $R_{\mathrm{ph}}\approx1$; the alloy-dominated regime has $R_{\mathrm{ph}}\ll1$. Strong alloy scattering, high carrier energy, and weak field collimation relative to angular diffusion all suppress the effective energy-loss rate. The product $A\thph^2$ is the rate at which the field repopulates the forward cone.

\prlhead{From phonon suppression to path-length narrowing}Cooling events during post-ionization re-acceleration reset or partially reset the carrier energy and broaden the ionization path length, so the relevant rate is $\GeffLO$, not $\GaLO$. Suppression narrows the path-length distribution in two ways: fewer POP events cool carriers below threshold during recovery, and the residual randomness becomes dominated by impact ionization itself rather than by repeated loss and re-acceleration cycles. Schematically,
\begin{equation}
P_{\mathrm{ion}}(x)\approx
\begin{cases}
0, & x<d,\\[2pt]
\lsp^{-1}\,e^{-(x-d)/\lsp}, & x\ge d,
\end{cases}
\label{eq:pathlength_schematic}
\end{equation}
with $\lsp/d\sim\Phi(R_{\mathrm{ph}},\Giv,\Gamma_{\mathrm{ii}},\Fld)$, where $\Gamma_{\mathrm{ii}}$ and $\Giv$ are the impact-ionization and intervalley rates. Evaluating $\Phi$ quantitatively requires device Monte Carlo, but the trend is robust: strong alloy scattering and weak effective POP cooling shorten $\lsp$ and reduce excess noise. Treating POP losses during recovery as a Poisson process gives a path-length spread $\sigma_x\approx(\hbar\wLO/q\Fld)\sqrt{d/\lLO}$ with $\lLO=v/\GeffLO$; identifying $1/\aEn\sim\sigma_x$ yields $\aEn\propto(\GeffLO)^{-1/2}$, so POP suppression raises the enabled ionization coefficient in the nonlocal regime (End Matter).

\begin{figure}[tbp]
\centering
\includegraphics[width=\columnwidth]{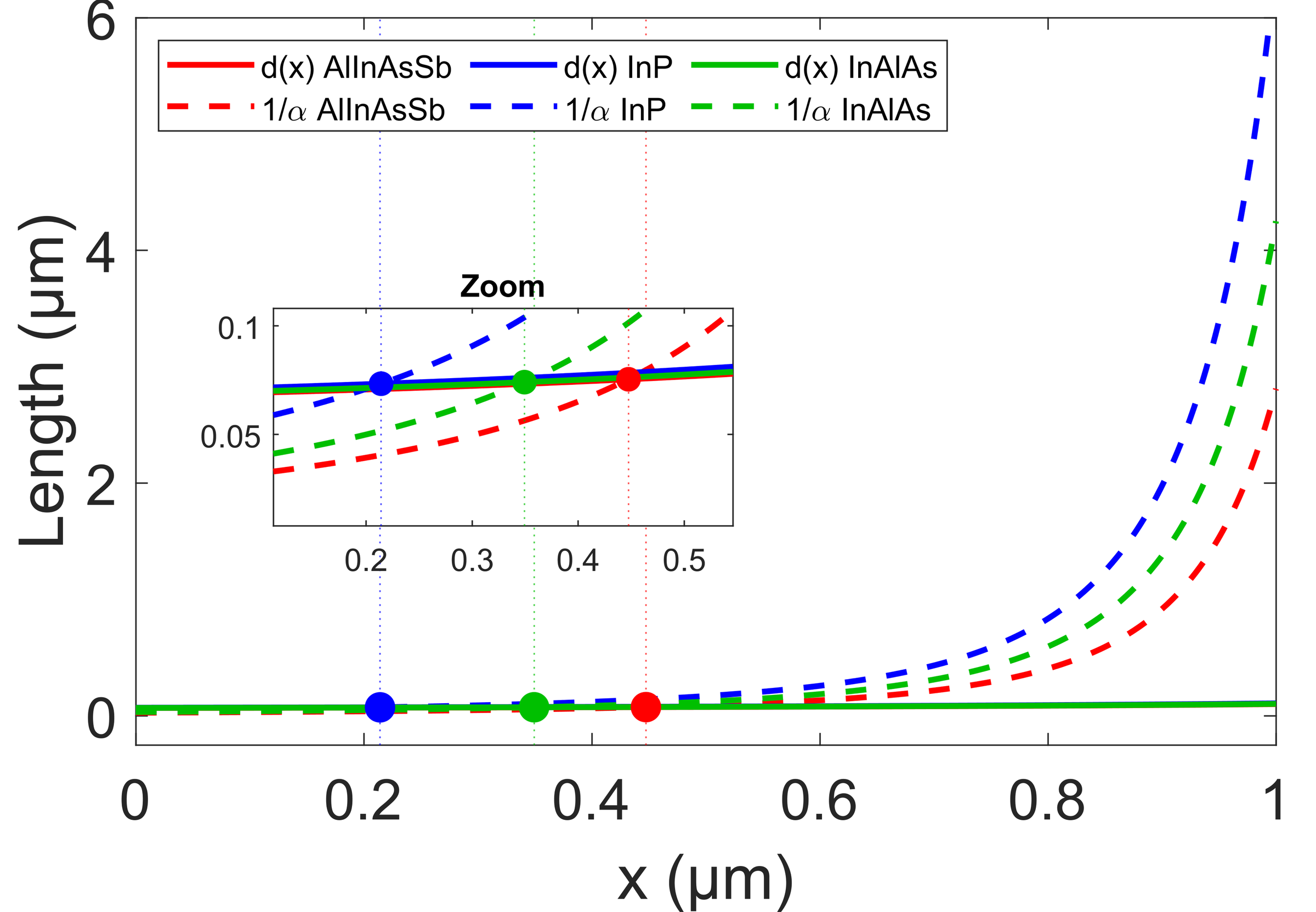}
\caption{Dead space $d$ (solid) and conventional local ionization length $1/\alpha$ (dashed) versus multiplication-region width $x$ at fixed gain for AlInAsSb (red), InP (blue), and InAlAs (green). Dots mark the crossover $d=1/\alpha$; for smaller $x$, $1/\alpha<d$, the local description fails, and multiplication is dead-space controlled and correlated. Inset: magnified crossover region.}
\label{fig:crossover}
\end{figure}

\prlhead{Correlation crossover and virtual staircase}Whether a device operates in the correlated regime of Eq.~\eqref{eq:centralcriterion} is set by two lengths that respond differently to the field. At fixed gain, a thinner multiplication region requires a higher field; $d=\ET/q\Fld$ then shrinks only in proportion to $1/\Fld$, whereas the conventional local ionization length $1/\alpha$ shrinks exponentially. Below a crossover width $1/\alpha$ therefore falls below the dead space (Fig.~\ref{fig:crossover}). Since $1/\alpha=d+1/\aEn$, no positive $\aEn$ can reproduce a local fit with $1/\alpha<d$: the local description is no longer self-consistent, ionization is controlled by the dead space, and $1/\aEn$ is small compared with $d$. The crossover lies at $x\approx0.21$, 0.35, and 0.45~$\mu$m for InP, InAlAs, and AlInAsSb, respectively: the alloy-disordered quaternary enters this correlated regime at the largest width, and hence the lowest field, as expected if alloy-suppressed POP cooling shortens the stochastic part of the path.

\begin{figure}[tbp]
\centering
\includegraphics[width=\columnwidth]{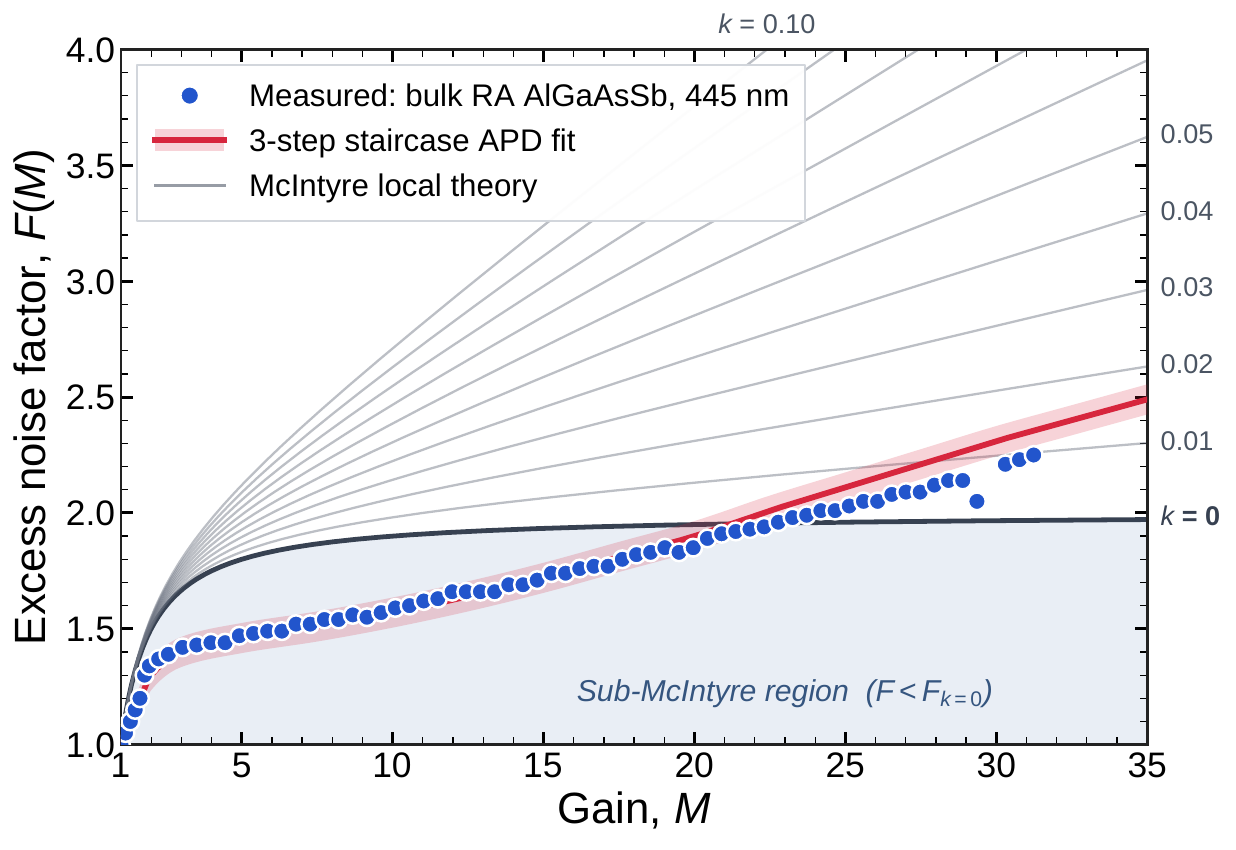}
\caption{Excess noise of a bulk random-alloy (RA) AlGaAsSb APD under 445-nm illumination (symbols; data from Ref.~\cite{Lewis2023}) compared with McIntyre local-field curves for $k=0$--0.10 (gray; $k=0$ dark) and a three-stage staircase-APD fit [red; Eq.~\eqref{eq:sapd_full} with $m=3$]. The shaded region $F<F_{k=0}$ is sub-McIntyre.}
\label{fig:sapd}
\end{figure}

In this regime successive ionizations act as the stages of a staircase APD~\cite{Capasso1983,March2021}. For a staircase with $m$ stages, per-stage ionization probability $P$, and hole-to-electron ratio $k_s$, $M$ and $F$ are known in closed form~\cite{Teich1986} [Eq.~\eqref{eq:sapd_full}]; for $k_s=0$,
\begin{equation}
\begin{aligned}
M&=(1+P)^m,\\
F&=1+\frac{1-P}{1+P}\left(1-\frac{1}{M}\right).
\end{aligned}
\label{eq:sapd_k0}
\end{equation}
Figure~\ref{fig:sapd} shows that this description, with $m$ as the only adjustable parameter ($k_s$ held fixed), reproduces the measured excess noise of a bulk random-alloy AlGaAsSb APD~\cite{Lewis2023} with $m=3$, including its sub-McIntyre segment and its crossing of the $k=0$ McIntyre curve near $M\approx20$, although the device contains no engineered steps. In the bulk device the stages are generated dynamically: each ionization generation spans on average $d+1/\aEn$, so we tentatively identify the effective stage number with the number of correlated generations across the multiplication width $W$,
\begin{equation}
m\simeq\frac{W}{d+1/\aEn}.
\label{eq:m_stages}
\end{equation}
In the correlated limit $1/\aEn\ll d$, $m\to W/d$ and $P\to1$, recovering a deterministic staircase with one ``step'' per dead space. The identification is a correlated-regime description: in the local limit $d\to0$ the stages become finer than the mean path and the discrete staircase picture no longer applies. Testing Eq.~\eqref{eq:m_stages} quantitatively requires $\aEn$ from device Monte Carlo.

\prlhead{Why Sb- and Hg-containing alloys}The mechanism requires strong elastic angular randomization and weak effective optical-phonon cooling. The quaternary disorder potential and alloy rate are given in Eqs.~\eqref{eq:quaternary_disorder} and \eqref{eq:alloy_rate}~\cite{Littlejohn1978,Shahnewaz2025,Hauser1976}. In Sb-based quaternaries, As--Sb sublattice mismatch and In-containing bonds increase the disorder strength~\cite{Shahnewaz2025}, consistent with alloy-dominated transport in Al$_{0.7}$InAsSb~\cite{Karimi2025}. HgCdTe multiplication is likewise affected by alloy scattering and dead space~\cite{Derelle2009} and exhibits low noise at high gain~\cite{Kinch2004}. The proposed mechanism places these systems in a common class of alloy-disordered materials with weak effective cooling and nonlocal multiplication.

\prlhead{Scattering hierarchy}Intervalley (IV) transfer can preserve or degrade the angular mechanism (Table~\ref{tab:hierarchy}). Alloy-dominated transport gives the lowest predicted noise; in the mixed regime, valley redistribution and energy loss can limit the benefit. When POP or IV scattering dominates, the statistics return toward McIntyre or dead-space-only behavior. The relevant hierarchy must therefore hold at the hot-carrier energies that control post-ionization recovery.

\begin{table}[tbp]
\caption{Predicted excess-noise trend for competing alloy, POP, and intervalley (IV) scattering. The alloy-dominated hierarchy satisfies Eq.~\eqref{eq:cubic_criterion} and yields the lowest $F(M)$.}
\label{tab:hierarchy}
\setlength{\tabcolsep}{3pt}
\begin{ruledtabular}
\begin{tabular}{p{0.2\columnwidth}p{0.35\columnwidth}p{0.35\columnwidth}}
Regime & Rate hierarchy & Predicted noise\\
\hline
Alloy dominated & $\Galloy>\GaLO>\Giv$ & Weak cooling; lowest $F(M)$, sub-McIntyre\\
Mixed alloy--IV & $\Galloy>\Giv>\GaLO$ & Residual IV loss; intermediate $F(M)$\\
POP dominated & $\GaLO>\Galloy,\Giv$ & Strong LO loss; high $F(M)$\\
IV dominated & $\Giv>\Galloy,\GaLO$ & Valley transfer; high $F(M)$\\
\end{tabular}
\end{ruledtabular}
\end{table}

\prlhead{Monte Carlo signatures}A three-dimensional $k$-space Monte Carlo model (End Matter) combines field streaming, forward-peaked POP emission, and elastic isotropic alloy scattering [Figs.~\ref{fig:kspace}(a) and \ref{fig:kspace}(b)]. POP scattering removes energy with little angular randomization; alloy events rotate the momentum and reduce cone occupation while the field streams the longitudinal component. The Monte Carlo model visualizes the angular mechanism rather than predicting absolute gain or noise; device Monte Carlo provides the quantitative comparison.

Device-level support comes from the multivalley Monte Carlo of Ref.~\cite{Karimi2025} for a 1~$\mu$m Al$_{0.7}$InAsSb APD calibrated against measured gain and noise. In this device simulation the fraction of electrons inside the $16^\circ$ cone falls from ${\approx}70\%$ with POP scattering alone to ${\approx}18\%$ when alloy scattering is included [Figs.~\ref{fig:kspace}(c) and \ref{fig:kspace}(d)], a direct confirmation of cone depopulation in a calibrated model. In Ref.~\cite{Karimi2025}, the measured excess noise was reproduced only when alloy scattering dominated optical-phonon scattering; with alloy scattering and impact ionization alone the simulated noise fell to HgCdTe-like levels, whereas optical-phonon-dominated transport raised it to near that of InP. The coefficient of variation of the ionization path-length distribution at $M\approx4$ was 0.46, 0.67, and 0.87 for HgCdTe, Al$_{0.7}$InAsSb, and InP, respectively~\cite{Karimi2025}---the ordering expected from Eqs.~\eqref{eq:Rph}--\eqref{eq:pathlength_schematic} if alloy-suppressed POP cooling narrows the spillover tail. The theory further predicts that, provided intervalley transfer does not dominate, the alloy-enabled excess-noise curve should fall below the local McIntyre $k=0$ curve over the gain range where dead-space correlation is strongest.

\prlhead{Design rule}Low-noise multiplication requires large $\Galloy$, small $\GeffLO$, electron-dominated ionization, and a width--field combination that enhances dead-space correlation, i.e., $\lsp/d\ll1$, with $\lsp$ decreasing as alloy suppression of POP cooling increases. In the alloy-broadened limit, a compact screening metric can be formed from the rate ratio and the POP cone scale,
\begin{equation}
\eta_{\mathrm{sub}}\equiv
\frac{\GaLO}{\Galloy}
\left(\frac{\hbar\wLO}{\ET}\right)^p,
\label{eq:eta}
\end{equation}
where $p$ depends on the assumed angular kernel and the energy dependence of the hot-carrier distribution. Combining the Poisson path-length estimate with Eq.~\eqref{eq:suppression_scaling} at $\Ekin\sim\ET$ (End Matter) gives $[1/(\aEn d)]^2\sim(\GaLO/\Galloy)(\hbar\wLO/\ET)^3$, i.e., $p=3$, and the material-level condition for sub-McIntyre multiplication is
\begin{equation}
\frac{\GaLO}{\Galloy}
\left(\frac{\hbar\wLO}{\ET}\right)^3\ll1,
\label{eq:cubic_criterion}
\end{equation}
up to order-unity factors. Equation~\eqref{eq:cubic_criterion} is a design heuristic rather than a universal threshold: it uses only bulk parameters, and quantitative prediction for a specific device still requires device Monte Carlo.

\prlhead{Experimental tests}The proposed mechanism can be tested by varying composition, temperature, width, and scattering channels. Composition changes the disorder potential and should favor lower noise where angular randomization is strongest, provided ionization remains electron dominated. Increasing phonon occupation should restore McIntyre-like behavior if cooling overtakes angular diffusion. Width--field sweeps should identify the minimum of $\lsp/d$ near the correlated regime in Fig.~\ref{fig:crossover}. Selective alloy, POP, and IV channel tests should connect the predicted noise ordering to transverse-momentum broadening and reduced cone occupation.

\prlhead{Discussion}Elastic angular randomization, usually regarded as mobility limiting, can benefit avalanche multiplication if it suppresses the cooling that broadens recovery lengths. The relevant metric is multiplication variance, not low-field mobility. Unlike engineered staircase APDs~\cite{Capasso1983}, the proposed bulk mechanism localizes ionization dynamically through dead-space recovery. Disorder is not unconditionally beneficial: excessive scattering can reduce velocity and heating, enhance intervalley transfer, or increase dark current. The useful regime preserves electron heating and electron-dominated ionization while reducing stochastic phonon loss. HgCdTe combines a small electron mass with strong electron--hole asymmetry; Sb-based quaternaries combine large disorder potentials with tunable band gaps.

\prlhead{Conclusion}We have presented a microscopic theory of correlated sub-McIntyre avalanche noise in alloy-disordered APDs based on the angular competition between forward-weighted POP emission and alloy-induced angular diffusion. Alloy scattering removes hot carriers from the POP emission cone, suppresses effective optical-phonon cooling, narrows the ionization path-length distribution, and promotes spatially correlated multiplication. This unifies the anomalously low noise of Sb- and Hg-containing APDs and indicates that APD noise can be engineered not only by reducing $k$ or inserting heterojunction steps, but by shaping the scattering hierarchy so that dead-space recovery dominates stochastic phonon-mediated spillover.

\label{prl:core-end}
\begin{acknowledgments}
The authors acknowledge discussions within the University of Virginia APD and transport theory groups. This work was supported by the National Science Foundation under Grant No.~\textbf{NSF 2430629}.
\end{acknowledgments}

\paragraph*{Data availability.}
The data that support the findings of this article are available from the authors upon reasonable request.

\bibliography{references}

\clearpage
\section*{End Matter}\label{prl:endmatter-start}
\appendix

\section{Fokker -- Planck steady state}\label{app:fp}
The continuity equation for the angular probability current on the sphere is
\begin{equation}
\begin{aligned}
\frac{\partial f}{\partial t}
&=-\frac{1}{\sin\theta}\frac{\partial}{\partial\theta}
\left(\sin\theta\,J_\theta\right),\\
J_\theta&=\dot\theta f-D_\theta\frac{\partial f}{\partial\theta}.
\end{aligned}
\end{equation}
Using $\dot{\theta}=-A\sin\theta$ gives Eq.~\eqref{eq:fp_full}. For $\theta\ll1$, $\sin\theta\simeq\theta$ and
\begin{equation}
\frac{\partial f}{\partial t}
=\frac{1}{\theta}\frac{\partial}{\partial\theta}
\left[A\theta^2 f+D_\theta\theta\frac{\partial f}{\partial\theta}\right].
\label{eq:fp_small}
\end{equation}
At steady state the integrated current must vanish for regularity at $\theta=0$, so $A\theta^2 f+D_\theta\theta\,\partial_\theta f=0$, i.e., $\partial_\theta\ln f=-(A/D_\theta)\theta$, which gives Eq.~\eqref{eq:theta0}. For a narrow distribution, the fraction inside a hard cone of half-angle $\thph$ is
\begin{equation}
P_{\mathrm{cone}}
=\frac{\int_0^{\thph} f\,2\pi\theta\,d\theta}{\int_0^\infty f\,2\pi\theta\,d\theta}
=1-\exp\!\left(-\frac{\thph^2}{2\thzero^2}\right),
\label{eq:pcone}
\end{equation}
which is Eq.~\eqref{eq:Rph}; for $\thzero\gg\thph$, $P_{\mathrm{cone}}\simeq\thph^2/2\thzero^2=A\thph^2/2D_\theta$, giving Eq.~\eqref{eq:suppression_scaling} with $D_\theta=C_a\Galloy$.

\section{POP emission-cone estimate}\label{app:cone}
From parabolic kinematics, $E'=\Ekin-\hbar\wLO$ gives $(k-k')/k=(k^2-k'^2)/[k(k+k')]\simeq\hbar\wLO/2\Ekin$, which is the denominator crossover angle $\phi_{\mathrm{peak}}$ in Eq.~\eqref{eq:smallangle_matrix}. The angular emission density, however, carries the phase-space factor $\sin\phi$,
\begin{equation}
\frac{d\Gamma}{d\phi}\propto\frac{\sin\phi}{q^2}\approx\frac{\phi}{(k-k')^2+k^2\phi^2},
\end{equation}
so the cumulative probability is only logarithmically concentrated near $\phi_{\mathrm{peak}}$. An order-one containment cone therefore scales parametrically as $\thph^2\sim\hbar\wLO/\Ekin$ [Eq.~\eqref{eq:phononcone}], up to order-unity factors. This square-root estimate should be read as a practical containment angle for forward-weighted emission, not as the literal denominator crossover.

\section{Derivation of the cubic criterion}\label{app:cubic}
If a carrier emits $N_{\mathrm{LO}}$ phonons while recovering, its path is lengthened by $\Delta x=N_{\mathrm{LO}}\hbar\wLO/q\Fld$, so $x=d+\Delta x$. Treating emissions as Poissonian, $\sigma_N^2\approx\langle N_{\mathrm{LO}}\rangle=\GeffLO d/v=d/\lLO$, with $\lLO=v/\GeffLO$ and $v=\sqrt{2\Ekin/\mstar}$, so
\begin{equation}
\sigma_x\approx\frac{\hbar\wLO}{q\Fld}\sqrt{\frac{d}{\lLO}}.
\label{eq:sigmax}
\end{equation}
Identifying $1/\aEn\sim\sigma_x$ (a transport-model identification rather than a strict identity) gives $\aEn\propto\lLO^{1/2}\propto(\GeffLO)^{-1/2}$. In the alloy-dominated limit, Eq.~\eqref{eq:suppression_scaling} gives $\GeffLO\sim\GaLO A\thph^2/\Galloy\sim(\GaLO/\Galloy)\,q\Fld\hbar\wLO/(\sqrt{2\mstar}\,\Ekin^{3/2})$; with $d=\ET/q\Fld$,
\begin{equation}
\frac{d}{\lLO}\sim\frac{\GaLO}{\Galloy}\,\frac{\hbar\wLO\,\ET}{2\Ekin^2}.
\end{equation}
Since $1/(\aEn d)\sim(\hbar\wLO/\ET)\sqrt{d/\lLO}$,
\begin{equation}
\left(\frac{1}{\aEn d}\right)^2
\sim\frac{\GaLO}{\Galloy}\,\frac{(\hbar\wLO)^3}{2\Ekin^2\ET}.
\end{equation}
Setting $\Ekin\sim\ET$ for the hot carriers that control ionization, the condition $1/\aEn\ll d$ becomes Eq.~\eqref{eq:cubic_criterion}, up to order-unity factors ($C_a$, the factor $1/2$ in $P_{\mathrm{cone}}$, and the cone-angle prefactor).

\section{Staircase APD expressions}\label{app:sapd}
For an $m$-stage staircase APD with per-stage electron ionization probability $P$ and hole-to-electron ratio $k_s$, the mean gain and electron-injection excess noise factor are~\cite{Teich1986}
\begin{subequations}\label{eq:sapd_full}
\begin{align}
M&=\frac{(1+P)^m(1-k_s)}{(1+k_sP)^{m+1}-k_s(1+P)^{m+1}},\\
F&=1+\frac{(1-1/M)(1-k_s)}{2+P(1+k_s)}
\Bigl\{-P+\frac{2(1-k_sP^2)}{1+k_sP}\nonumber\\
&\qquad\times\Bigl[\frac{Mk_s(1+P)}{1-k_s}+\frac{1}{1+P}\Bigr]\Bigr\}.
\end{align}
\end{subequations}
For given $m$ and $k_s$, $P$ is eliminated by inverting the first expression at each $M$, so $F(M)$ follows with no further parameters. Setting $k_s=0$ gives Eq.~\eqref{eq:sapd_k0}; for $k_s=0$ the gain is capped at $2^m$, while $k_s>0$ allows larger gain through hole feedback.

\label{prl:endmatter-end}
\end{document}